\documentclass[twocolumns]{aastex631}

\usepackage{newtxtext,newtxmath}
\usepackage{graphicx}
\usepackage{xcolor}
\usepackage{subfigure}
\usepackage{ulem}
\usepackage{mathtools}
\usepackage{amsmath}
\usepackage{lastpage}

\def\me{$\,{\rm M}_{\oplus}\,$}

\def\h2o{H$_2$O}
\def\sio2{SiO$_2$}
\def\gc3{g\,cm$^{-3}$}

\shorttitle{Interiors of ice-rich planets}
\shortauthors{Vazan et al.}

\begin{document}

\title{A new perspective on interiors of ice-rich planets:\\
Ice-rock mixture instead of ice on top of rock}

\author{Allona Vazan}
\affiliation{Astrophysics Research Center of the Open university (ARCO), The Open University of Israel, 4353701 Raanana, Israel}
\altaffiliation{Department for Natural Sciences, The Open University of Israel}
\email{vazan@openu.ac.il}

\author{Re'em Sari}
\affiliation{Racah Institute of Physics, The Hebrew University of Jerusalem, Jerusalem 91904, Israel}

\author{Ronit Kessel}
\affiliation{Institute of Earth Sciences, The Hebrew University of Jerusalem, Jerusalem 91904, Israel}

\begin{abstract}
Ice-rich planets are formed exterior to the water iceline and thus are expected to contain a substantial amount of ices. The high ice content leads to unique conditions in the interior, under which the structure of a planet is affected by ice interaction with other metals. We apply experimental data of ice-rock interaction at high pressure, and calculate detailed thermal evolution for possible interior configurations of ice-rich planets, in the mass range of super-Earth to Neptunes (5-15\me). We model the effect of migration inward on the ice-rich interior by including the influences of stellar flux and envelope mass loss. 
We find that ice and rock are expected to remain mixed, due to miscibility at high pressure, in substantial parts of the planetary interior for billions of years. We also find that the deep interior of planetary twins that have migrated to different distances from the star are usually similar, if no mass loss occurs. Significant mass loss results in separation of the water from the rock on the surface and emergence of a volatile atmosphere of less than 1\% of the planet's mass. The mass of the atmosphere of water / steam is limited by the ice-rock interaction.
We conclude that when ice is abundant in planetary interiors the planet structure may differ significantly from the standard layered structure of a water shell on top of a rocky core. Similar structure is expected in both close-in and further-out planets. 
\end{abstract}


\section{introduction}\label{sec:intro}
Planets form by accretion of solids and gas from the protoplanetary disk, and their composition depends on the formation location in the disk. Ice-rich planets were formed exterior to the water iceline\footnote{The icelines of other volatiles are located exterior to the water iceline. Ice-rich planets that form further out also contain other ices, but water is expected to be the most abundant.}, where water ice is as abundant as rock \citep{lodders03}. Thus, they contain a substantial fraction of ice. Based on formation models, these planets didn't reach the runaway gas accretion phase to become gas giant planets, but usually had conditions to accrete some fraction of gaseous envelopes \citep{mordasini12,lambrechts14,morbi15}.
The interior structure of these planets is poorly understood. 
The interior of ice rich planets is usually modeled as a 4 layer object: iron core overlaid by a rocky mantle, surrounded by an icy shell and in some cases covered with a gaseous (usually hydrogen and helium) envelope \citep[e.g.,][]{rogersseag10,rogers11,noack16,dorn17}. While this structure is assumed for simplicity based on density consideration, it should be reexamined for two main reasons:

(1) In the basic Core Accretion model, the accreted solids (metals) form a solid core, surrounded by a solar composition gaseous envelope \citep{pollack96,alibert05a}.
But detailed calculations of material and energy deposition during solid accretion indicate ablation of solids in the gaseous envelope, and emergence of gradual composition distribution, where the metal mass fraction decreases outward in the planet interior \citep{lozovs17,brouwers18,boden18,helledsteven17,valletta19,valletta20,ormel21}. 
In many cases most of the rock in the planet is in a vapor form \citep{boden18,ormel21}.
Gradual composition distribution in the interior is also consistent with measurements of {Jupiter and Saturn \citep{lecontechab13,fuller14,vazan16,wahl17,vazan18b,debras19} and Uranus and Neptune \citep{marley95,podolak00,helled11UN,vazanhel20} in} our solar system. Thus, planets {that form} with gas envelopes may have a gradual metal distribution in the interior, in opposed to a few distinct shell structures. 

(2) The differentiation of an initially mixed\footnote{Here we assume the ice and rock to be initially mixed. Some formation scenarios may separate or partially separate the accreted ice and rock, as ice evaporates at lower temperatures than rock. See Section~\ref{sec:mod_init} for details and Section~\ref{sec:dsc_cvt} for discussion.} planet interior into distinct layers by gravity occurs only if the interior material is {chemically} demixed (immiscible). Chemical mixing / demixing is determined by the thermodynamic properties of the components, namely, their tendency to interact as a function of pressure and temperature \citep{stevenson13}.
Laboratory experiments of ice-rock interaction in the context of Earth composition have been conducted for many years \citep[e.g.,][]{kennedy62,shenkep97,schmidt98,kessel15}.
These laboratory experiments show that ice and rock are miscible in each other at high (GPa) pressure.
Recent laboratory experiments and theoretical work at tens and hundreds of GPa pressure support this finding \citep{nisr20,kim21,gao21}
Yet, no attempt was made to link these experimental results to long term thermal evolution of ice-rich planets.
Several works suggested that ice and rock mixtures are consistent with Uranus and Neptune measured properties \citep{marley95,podolak00,vazanhel20,bailey21}. However, these suggestions were not justified by material interaction properties or thermodynamic behavior, and were meant to fit or explain observational data.

Although ice-rich planets formed exterior to the water iceline, common planet-disk interaction theories predict planetary migration in the protoplanetary disk to be pervasive \citep{ward97,papaloiz02,bitsch13,mordasini18}. 
Observational evidence of an evaporating or disrupted ice-rich planet around white dwarfs \citep{veras20} and of water in an  atmosphere of a close-in Neptune \citep{kreidberg20} indicate that ice-rich planets may get close-in and maintain their ice.
When a planet migrates inward in the protoplanetary disk, its interior may change by the higher stellar flux over billions of years. The stellar flux reduces the planet luminosity and thus slows down the planet interior cooling. In addition, the higher stellar flux absorbed by the outer layers of the gas envelope can stimulate mass loss processes. 
Both effects can affect the thermal evolution of the interior. 

Here, we explore possible interior configurations of ice-rich planets. We include experimental results of ice-rock interaction, possible initial structures, and migration effects. In section~\ref{sec:mod} we outline our model and the recent knowledge of interior formation, {ice-rock} interaction, and non-adiabatic thermal evolution. In section~\ref{sec:rslt} we present resulting interior structures of planets with and without mass loss. We discuss implications and caveats in section~\ref{sec:dsc}, and summarize our conclusions in section~\ref{sec:cncl}.

\section{Model}\label{sec:mod}
\subsection{Initial structure}\label{sec:mod_init}

Exterior to the water iceline rocks and ice are {both available} as solids. Based on the outer solar system composition water is the most abundant volatile, and it is about half of the solid mass exterior to the water iceline \citep{lodders03}. Although ice to rock mass ratio may vary with disk parameters (see discussion in Section~\ref{sec:dsc_unc}), in this study we take the solar system ratio where the solid building blocks outside the iceline are composed of 50\% water ice and 50\% rock by mass. 
It is unclear whether the ice and rock stay mixed during the accretion process. Some models of solid accretion suggest evaporation of the ices in the envelope, while the rocks accreted to form the core \citep{podolak88,horiikoma11,venturini16}. 
However, later studies show that rocks also dissolve in the deep envelope of the growing planet, already when the planet reaches a mass of a few Earth masses \citep{boden18,brouwers18,valletta19,brouwers20}. 
Although ice vaporizes at higher altitudes, when both ice and rock vaporize in the envelope they may tend to mix \citep{lock17}. 
In addition, large accreted solids that reach the deep interior without breaking (e.g., km size planetesimals), keep the ice to rock ratio in the deep interior similar to the ratio in the disk. 
Clearly, the initial ice and rock distribution in the interior requires further studies, but formation studies hint that some fractions of ice and rock may be mixed in the planet interior at the end of the formation phase.
For simplicity, and as a first step, we assume here that all of the ice and rock are initially mixed in the interior.

Metal-dominated planets have less gas than metals, since if the gas to metal mass ratio exceeds unity the planet reaches the runaway gas accretion phase and becomes a gas giant \citep[e.g.,][]{pollack96,brouwers20}\footnote{The {\it critical core mass}, which is the minimum core mass that is required to initiate rapid gas accretion, is replaced by {\it critical metal mass}, which is a wider definition that also includes planets with polluted envelopes and gradual metal distribution \citep{brouwers20,ormel21}.}. The amount of accreted gas depends on parameters such as radiative opacity, accretion rate, and distance from the star \citep{lammer14}.
The radius-mass relation of observed sub-Neptunes predicts a gas fraction of up to 10\% in mass \citep{lopezfor14}, consistent with theoretical studies \citep{ikoma12,bitsch15,johansen17}.
The ice-rich candidates in our solar system, Uranus and Neptune, have about 10-20\% of gas in mass \citep{helled11UN,nettel13}. 
Thus, we take 10\% hydrogen and helium (H, He) in solar ratio to be our standard initial gas fraction. As will be shown below, the exact fraction of gas doesn't change the conclusions we make in this work about the interior structure of ice-rich planets.

\begin{figure}
\centerline{\includegraphics[width=9.5cm]{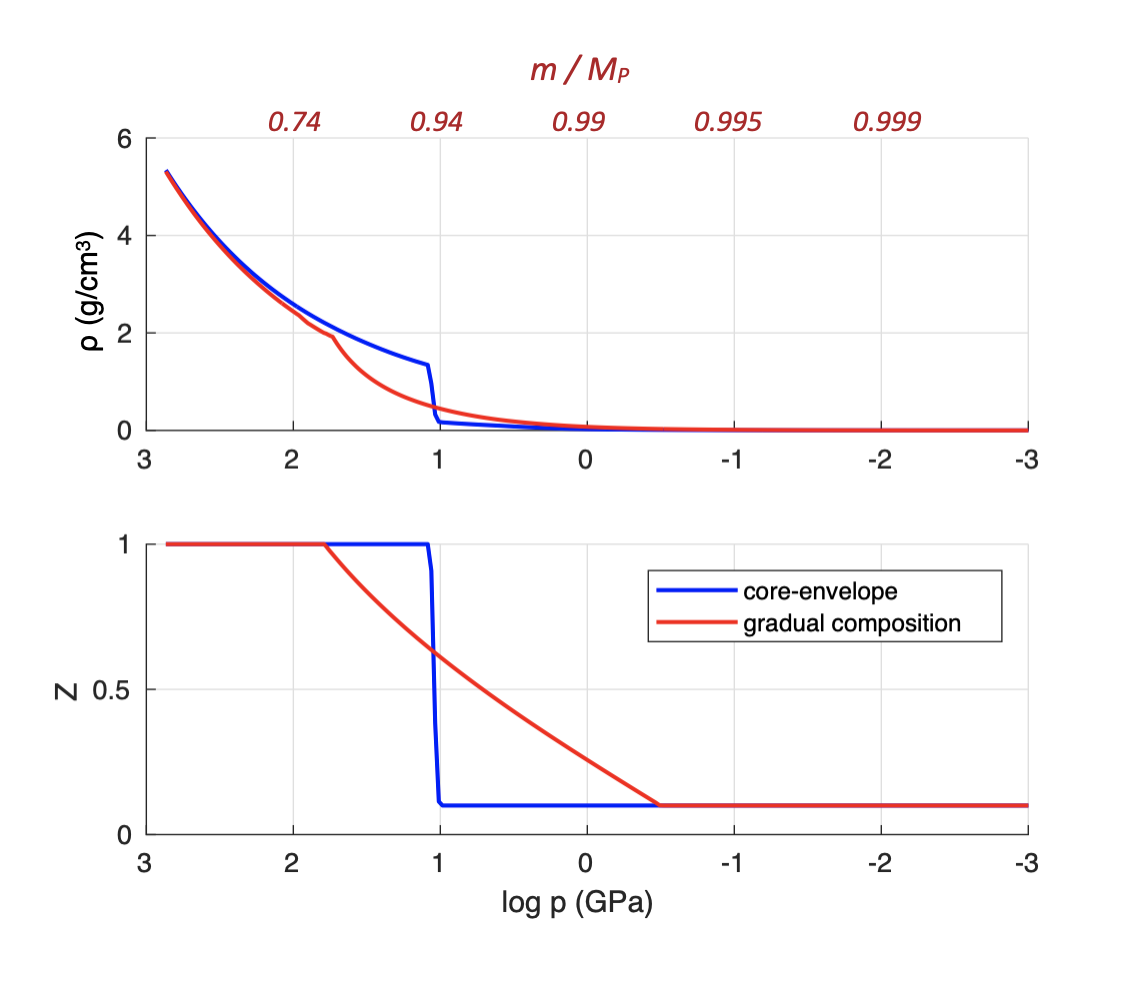}}
\caption{Initial density profile (top) and metal (ice+rock) distribution (bottom) of two representative models: core-envelope structure (blue) and gradual composition distribution (red). The modeled planets are 15\me where 90\% of the mass is ice and rock in equal mass fractions (1:1 ratio), and 10\% of the mass is hydrogen and helium in solar ratio. The planets are located at 10AU from a sun-like star. The upper red axis indicates the mass fraction located deeper than this pressure for the gradual structure (the core-envelope case varies by less than 1\%).}\label{fig:initial}
\end{figure}

We consider two different initial structures: gradual metal distribution and core-envelope structure. 
The gradual structure, motivated by new planet formation models \citep{boden18,valletta20,ormel21}, has a deep pure metal region, followed by a gradual decrease in {metal mass fraction (Z)} down to Z=0.1 in the outer envelope. 
For comparison we consider also the same composition in a simple core-envelope structure, with a pure Z core surrounded by a H,He envelope with uniform Z=0.1 enrichment.
In Fig.~\ref{fig:initial} we show the initial density and metal distribution for 15\me planets of these two models.

To summarize, our initial standard models are of gradual composition structure and of core-envelope structure. For clarity, we focus our examples in this paper on  5-15\me planets, initially composed of 90\% metals (ice and rock in 1:1 {mass} ratio), and 10\% H,He (solar ratio). 

\subsection{Ice-rock interaction}\label{sec:mod_mix}

After formation, the long term ice and rock distribution in the interior depends on the tendency of ice and rock to mix or demix in different pressure-temperature regimes \citep{stevenson13,stevenson14}. 
Chemical separation (demixing) of ice and rock leads to a differentiation into different compositional layers by the planet's gravity field, based on their densities. 
If, on the other hand, ice and rock are chemically mixed (miscible), then no differentiation into layers takes place. 
At high enough pressure and temperature the ice-rock mixture reaches the second critical end-point (SCP). At pressures or temperatures above the SCP the rock and ice are expected to be completely miscible in each other, in a single supercritical fluid phase.
Below the SCP ice and rock can be partially miscible or immiscible (demixed) as a function of the ice to rock ratio. {More details about the critical point are provided in appendix~\ref{apx:scp}}.

The SCP of a mixture is difficult to predict theoretically, and is determined by laboratory experiments.  
In such experiments the interaction of ice and rock up to a pressure of a few GPa is examined. The experimental results are function of pressure and temperature\footnote{The results are also function of ice to rock mass ratio (see appendices~\ref{apx:curve}), and of rock type (see appendix~\ref{apx:rock}).}. For a given ice to rock mass ratio and a given rock type, two curves in the pressure-temperature space characterize the mixture interaction: the melting curve (wet solidus) below which the ice-rock separation is complete, and the miscibility curve above which the chemical mixing is complete.
In order to apply this knowledge to planetary interiors, we first collect the experimental data, as a function of pressure and temperature.
More details on the derivation of the wet solidus and miscibility curve, and how they are linked to the SCP are provided in appendix~\ref{apx:curve}.

We use data from laboratory experiments at high pressure by \cite{kessel15,melekhova07} and \cite{grove06} for the interaction of water-ice and Earth-like peridotitic rock\footnote{Peridotite is the dominant rock in the upper mantle of the Earth, and representative of solar system refractory composition. More about the choice of peridotite in Appendix~\ref{apx:rock}.}. 
In Fig.~\ref{fig:PT_space} we present the resulting map of mixing and demixing of ice and rock: gray shaded area signifies where ice and rock are completely mixed (miscible), the brown shaded area is where ice and rock are partially mixed, and in the white area ice and rock are demixed (immiscible, separated). 
The water vapor curve given in \cite{alduchov96} (dashed black) indicates the liquid-gas phase transition of pure water, up to the first critical end-point, which is the critical point of pure water (see appendix~\ref{apx:scp}).

The picture we draw in Fig.~\ref{fig:PT_space}, and in particular the boundaries between areas, is an example of peridotitic rock and water in a 1:1 mass ratio. 
The exact values and boundaries may vary with rock type (appendix~\ref{apx:rock}), and water fraction (appendix~\ref{apx:curve}). 
However, those uncertainties, in the range of one GPa and of a hundred Kelvin, are quite small from the perspective of the entire pressure-temperature range of super-Earth to Neptune interiors.
More about the limitations and caveats of the model appears in Section~\ref{sec:dsc} and the appendices.

For reference we plot in Fig.~\ref{fig:PT_space} the pressure temperature { range} of the Earth \citep{pearson03,nomura14}, and possible structure of Jupiter \citep{vazan18b}. 
As shown, the Earth and Jupiter are mainly in the complete ice-rock mixture regime. However, while Jupiter probably formed the exterior of the ice-line and thus is ice-rich, Earth is ice-poor. If the Earth were formed in the ice-rich region of the protoplanetary disk\footnote{The data in Fig.~\ref{fig:PT_space} cannot be applied to the real dry Earth, because dry and wet rocks differ in their properties. The Earth curves indicate that an ice-rich Earth-like planet would have mixed ice-rock in its interior.}, the ice would have been mixed with the rock in its interior, probably also preventing the differentiation of the iron core (see the next paragraph).
Moons, on the other hand, are expected to be differentiated. For example, the large moon of Jupiter - Ganymede - has a central pressure of about 10\,GPa and central temperature of about 1500K \citep{shibazaki11}, already below the miscibility regions. 

{\bf Iron-ice-rock interaction} The formation of Fe-dominated metallic cores{, as exists in the centre of the Earth,} is a consequence of an insufficient O/Fe ratio \citep{tronnes19}. The fraction of oxygen in ice-rich planets is much larger than in dry planets, and the iron is expected to be oxidized and therefore remain in the rock rather than segregate as metallic core alloys, while the leftover hydrogen dissolves in the rock or escapes by outgassing. 
As a result, the schematic picture of mixed metals in the interior of ice-rich planets may hold also for iron. 
Nevertheless, this estimate is based on limited knowledge. Experiments of ice-rock-iron mixtures in ice-rich conditions are lacking, and further studies are required to constrain its properties in planetary conditions. 

\begin{figure}
\centerline{\includegraphics[width=10cm]{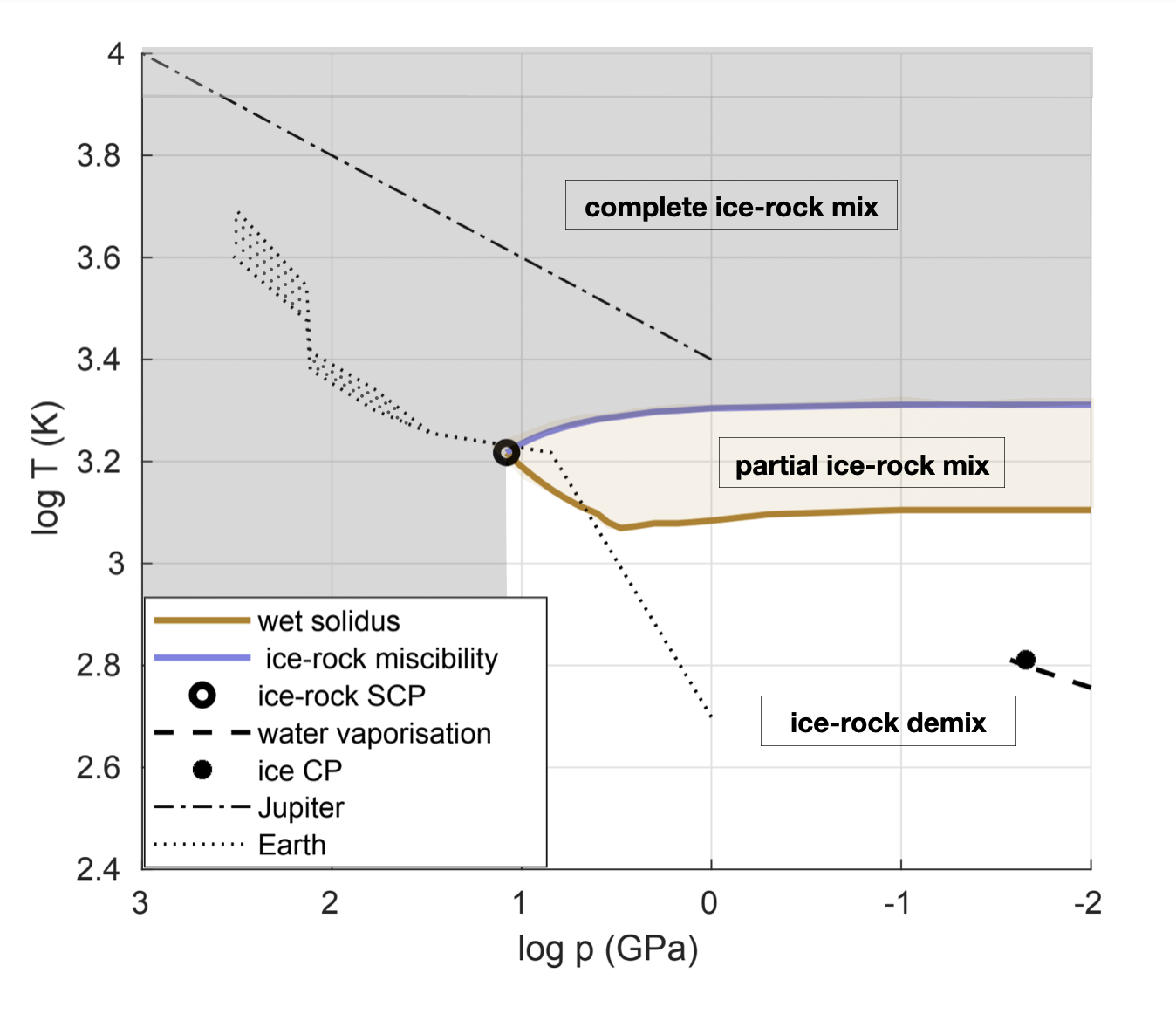}}
\caption{Ice-rock interaction regimes in pressure-temperature space, based on laboratory experiments data of Kessel et al. 2015, Melekhova et al. 2007, and Grove et al. 2006. Ice-rock miscibility curve (blue solid line) and wet solidus (brown solid line) represent Earth-like rock and water ice in 1:1 mass ratio. These lines and the ice-rock second critical point (black circle) define three regions in the $(p,T)$ plane: complete ice-rock mix (gray shaded area), partial ice-rock mix (brown shaded area), and complete immiscibility (white area). The water vaporization curve signifies pure water phase transition (dashed black){, up to the water critical point (black dot)}.
Pressure-temperature profile in the interiors of the Earth (dotted black range) and Jupiter (dashed dotted black) are shown for reference.}\label{fig:PT_space}
\end{figure}

\subsection{Thermal and structural evolution}\label{sec:mod_therm}

Next, we apply the data from Fig.~\ref{fig:PT_space} to the thermal evolution of ice-rich planets. 
Thermal evolution determines the planet temperature profile in time; the simplest approach for thermal evolution is of adiabatic (isentropic) planets. While this approach can fit to some level a structure of 2-3 uniform shells, it cannot be used to model gradual composition distribution in planetary interiors because it disregards the mutual effect of composition distribution on the heat transport \citep{ledoux47,lecontechab12,vazan15,muller20}. These effects can also be significant in the interiors of ice-rich planets, as shown in \cite{vazanhel20} for the interior of Uranus.

Our thermal evolution model is based on \cite{vazan18b} with modifications to {metal-dominated} sub-Neptune planets, as described in \cite{vazan18c}. The model allows for heat transport by convection, radiation, and conduction, depends on the local conditions in time. The model encapsulates the mutual effect of composition distribution on the heat transport, and the non-adiabatic thermal evolution when required.
In convective regions we calculate the change in structure by convective mixing, according to the mixing length recipe \citep[see][]{vazan15}. 

We use equation of state (EoS) for a mixture of ice, rock, hydrogen and helium (improved version of \cite{vazan13}), based on the additive volume law. For simplicity, we first calculate an EoS for a mixture of ice and rock. Our rock and ice EoS includes a liquid-solid phase and a vapor phase, and thus can also be used to model steam atmospheres. We calculate the mixtures of ice, rock, hydrogen and helium in the evolving structure self-consistently during the thermal evolution. 
For radiative opacity we use the method of \cite{valencia13} to fit the  \cite{freedman08} tabular opacity for planetary atmospheres. The metal fraction in the opacity calculation is linked to the outer envelope metallicity in the model. Radioactive heating by long term radioactive elements in the rock, and heat from solid contraction are also included in the model, as described in \cite{vazan18c}.

\subsection{Migration and mass loss}\label{sec:mod_loss}

The stellar flux is included as a boundary condition of the interior model. The optically thin outer atmosphere is assumed to be gray and plane parallel \citep{vazan18c}.
We use a simplified model of linear migration, where we change the location (stellar flux) from formation location to final location within 3 Myr. {The migration timescale (Myr) is much shorter than the evolution timescale (Gyr), and therefore the exact migration path is negligible for the aims of this work}.

As the planet migrates close enough to the star, the high stellar flux may cause mass loss. 
Mass loss from migrated-in planets can be via several mechanisms, among them photoevaporation \citep{owenwu13,owenwu17}, hydrodynamic wind \citep{ginzburg16}, Jeans escape \citep{lammer03,zeng19}, and photodissociation \citep{howe20}.
Current mass loss predictions {for metal-dominated planets} vary with uncertain stellar properties and timescales  \citep[e.g.,][]{Lopez12,mordasini20,king21}.
Metal-rich envelopes add further complication to the mass loss calculation by affecting the planet radius, atmospheric composition, and mean molecular weight \citep{venturini20a,malsky20}. 
As of today, theories for mass loss from metal-rich envelopes are lacking and fail to explain some of the observations \citep{kasper20}.
Since in this work we focus on the effect of mass loss on the interior structure, and not on the mass loss process itself, we consider here 3 simple edge scenarios for close-in planets{, by different studies}: no mass loss {(mainly for comparison)}, loss of the hydrogen {\citep[e.g.,][]{hu15})}, and loss of {all the hydrogen and helium \citep[e.g.,][]{zeng19}}.

\section{Results}\label{sec:rslt}

The results are shown for planetary ages of 10 Gyr. Earlier in the evolution the temperatures are higher and therefore thermal differences between models are smaller. In addition, at higher temperature more mass is in the ice-rock mixed area, as shown in Fig.~\ref{fig:PT_space}. Therefore, the models at the age of 10 Gyr are the lower bound for the effects we study here.

\subsection{Interiors of different structure and location}\label{sec:rslt1}

\begin{figure}
\centerline{\includegraphics[width=9.5cm]{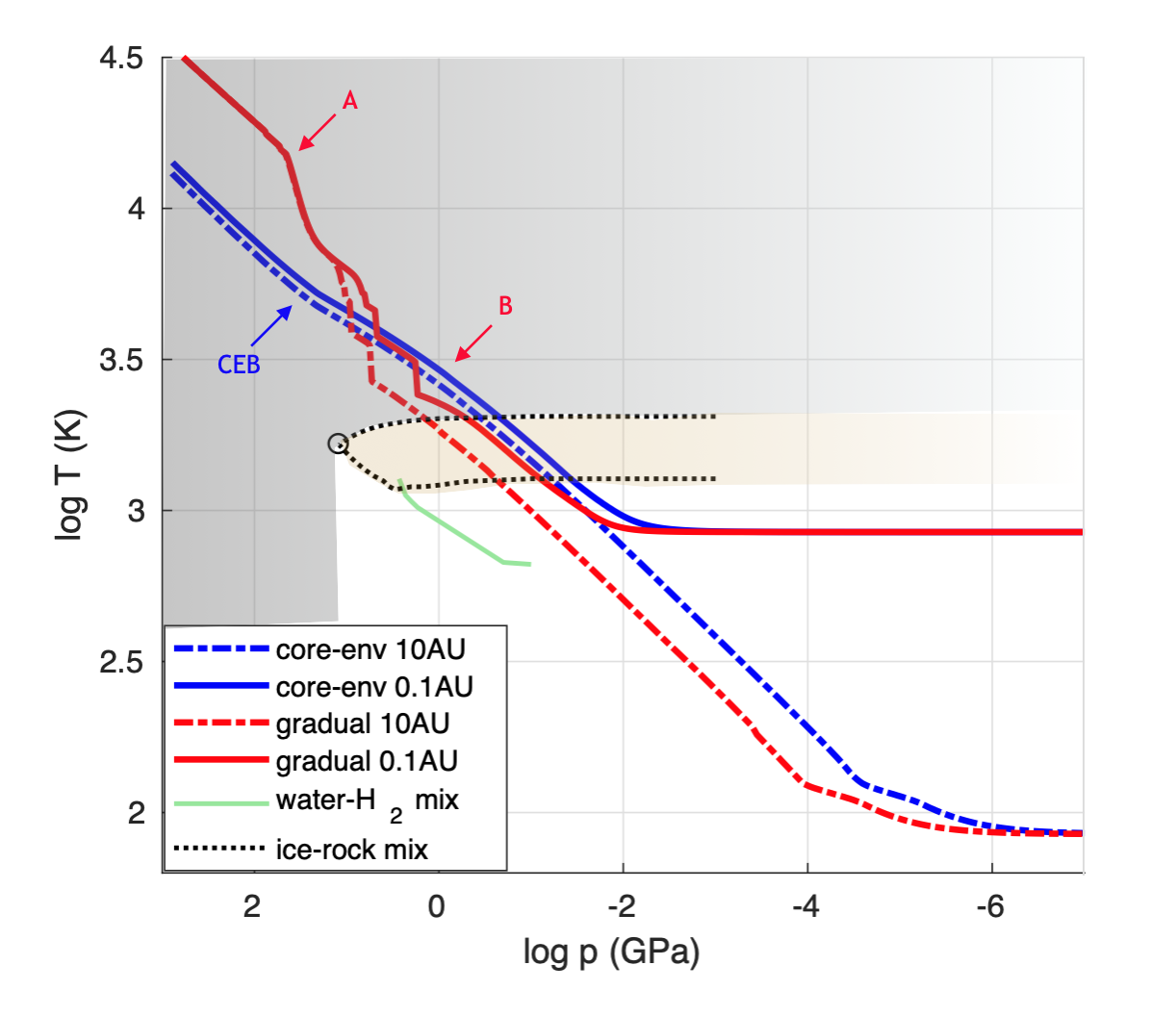}}
\caption{Pressure-temperature profiles of 15\me planets with 10\% gas at the age of 10 Gyr. The same metal (ice and rock) content is gradually distributed (red) or located in a pure metal core surrounded by a gas envelope (blue). Core-envelope boundary (CEB) is marked with the blue arrow, and the gradual composition distribution region expands between the red A and B arrows. The planets are located at 10AU (dashed) or migrated to 0.1AU (solid). In the shaded areas, taken from Fig.~\ref{fig:PT_space}, ice and rock are expected to stay mixed. Above the hydrogen-water mix curve (green), water is miscible in hydrogen. The mass in the shaded area is $>$99\% of the planet's mass for all cases.}\label{fig:PT_0110}
\end{figure}

In Fig.~\ref{fig:PT_0110} we show two types of interior structures of 15\me ice-rich giants, as evolved from the structures in Fig.~\ref{fig:initial}: core-envelope (blue) and gradual Z distribution (red). The initial energy content and composition mass fraction of all cases are similar. 
The difference between the temperature profile of the core-envelope structure and the gradual Z structure on Gyrs time is due to the composition distribution effect on the heat transport. The composition gradient suppresses convection and therefore the heat flux outward is lower. As a result, the deep interior of the gradual Z case is much hotter, while the outer layers cool more efficiently.
We also find that the gradual structure maintains the composition gradient from formation, with minor redistribution by convective-mixing, mainly in the outermost convective zone. This result is anticipated for a steep composition gradient, which is more difficult to mix \cite[e.g.][]{vazan16}.
Since {metal-dominated} interiors must have steep metal gradients (shallow metal gradients require large gas fraction) we find that convective-mixing is insignificant in planets with {metal-dominated} interior. 

On the background of the planet profiles in Fig.~\ref{fig:PT_0110} we show the ice-rock mixing and demixing areas from Fig.~\ref{fig:PT_space}. 
We find that the mixed ice-rock regime covers {a large mass fraction} of the planet in all cases, where the metal-rich interior (marked in arrows) is clearly within the ice-rock miscibility region. 
We also plot in Fig.~\ref{fig:PT_0110} the water-hydrogen miscibility curve of \cite{bali13}. Above the water-hydrogen miscibility (green curve) the water is mixed with the hydrogen. 
{Hence, the water in the envelope region between about 700\,K and 1300\,K is immiscible in rock and miscible in hydrogen\footnote{Above about 1300\,K ice and rock are miscible in each other, and hydrogen is expected to also be partially miscible in water and molten rock \citep{kite19}}.}

We also compare in Fig.~\ref{fig:PT_0110} similar planets at different distances from the star. The planets have the same initial properties, one is located at 10AU (dashed) and one migrated to 0.1AU (solid) from the star. No mass loss is considered here, to isolate the effect of stellar irradiation on the cooling. As can be seen, even after 10 Gyr of evolution the deep interior is only slightly affected by the stellar irradiation, especially if the planet composition is gradually distributed. Although the outermost layers of the closer-in planet expand by the stellar flux, causing a significant radius {inflation}, these layers contain little mass, and the deep interior is almost not affected. Unlike in gaseous planets, the radius (volume) of the metal-rich deep interior isn't very sensitive to its energy content \citep{seager07,valencia06,rogers11} and the structure remains similar. 
Thus, the deep interior of twin {Neptune-like planets that migrated to different distances from the star (here 10AU and 0.1AU) is very similar, if no mass loss takes place.}

\subsection{Interiors after mass loss}\label{sec:rslt2}

As the planet losses its lighter materials by the stellar flux, the mean molecular weight of the outermost layer increases. 
Once the outer layers become oversaturated the excess of metals will rainout to deeper layers, and enrich them with metals. In the case of gradual composition distribution this process slowly flattens the interior composition gradient outside-in, allowing for more rapid cooling by large scale convection in the forming metal-rich outer envelope. 

\begin{figure}
\centerline{\includegraphics[width=9cm]{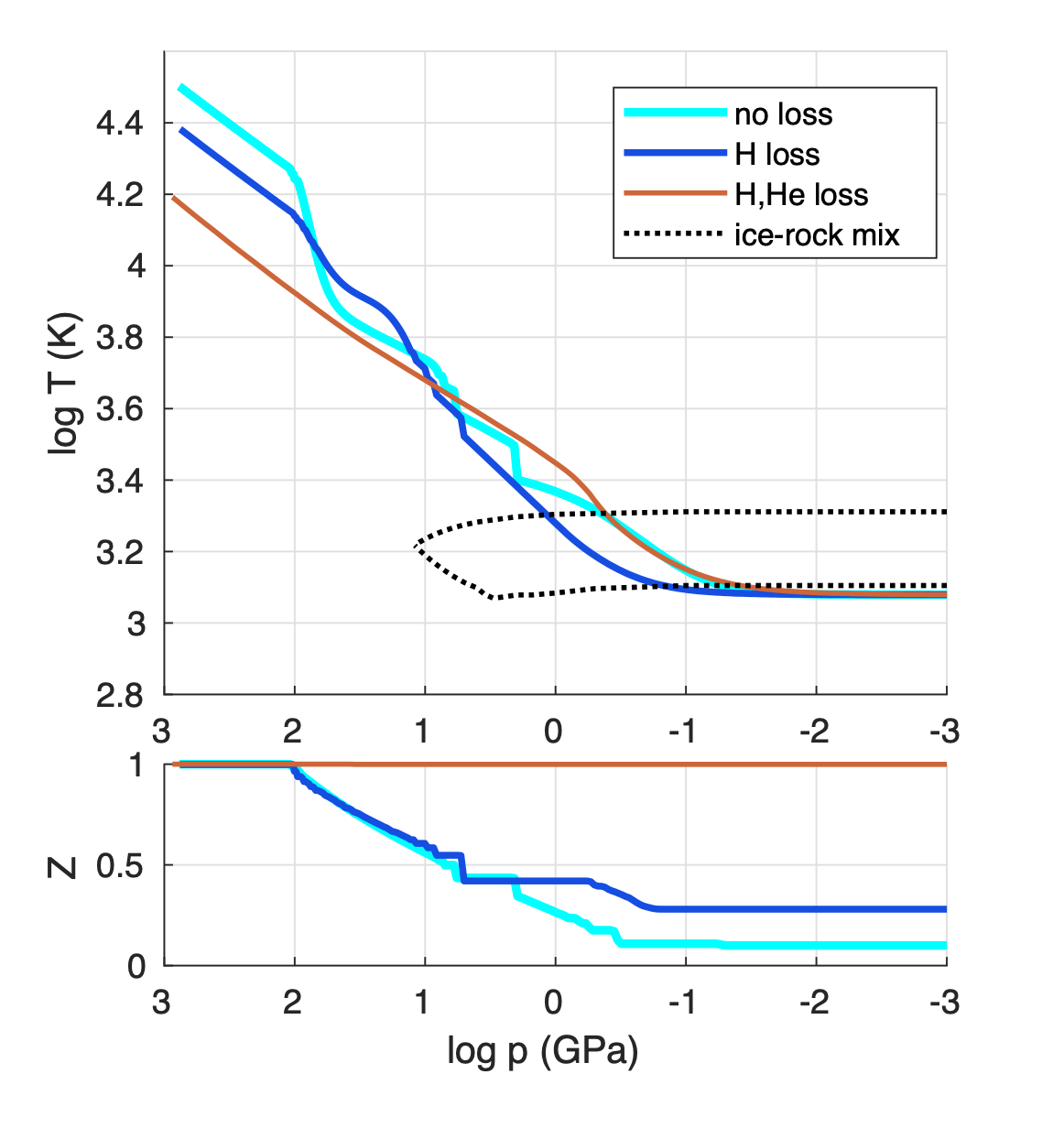}}
\caption{Temperature profile (top) and metal distribution (bottom) vs. pressure for initially 15\me planets that have migrated to 0.05AU, with different mass loss scenarios: no loss (cyan), loss of hydrogen (blue) and loss of all gas (brown). The ice-rock mixing curves (dotted-black) are taken from Fig.~\ref{fig:PT_space}.}\label{fig:PT_005}
\end{figure}

In the upper panel of Fig.~\ref{fig:PT_005} we show the pressure-temperature profile of initially 15\me planet with gradual composition distribution that migrated to 0.05AU, under 3 different mass loss assumptions: no mass loss (cyan), mass loss of all hydrogen (blue), and mass loss of all hydrogen and helium (brown). 
We find that the completely demixed ice and rock outer region has a mass fraction $<$0.01\% for the cases shown in this figure. 
Namely, even without gaseous envelope most of the ice and rock in the interior are { above the miscibility (mixing) curve}. 
{The faster (adiabatic) cooling of the gas-free planet by large scale convection, in contrast to the slower cooling of planets with gradual composition distribution, reduces the temperatures in the deep interior.}
In the bottom panel, we show the metal (ice+rock in a 1:1 mass ratio) distribution in the interior for the 3 cases. The metal distribution is affected by the loss of light materials and by the moderate convective-mixing. 

\begin{figure}
\centerline{\includegraphics[width=9.5cm]{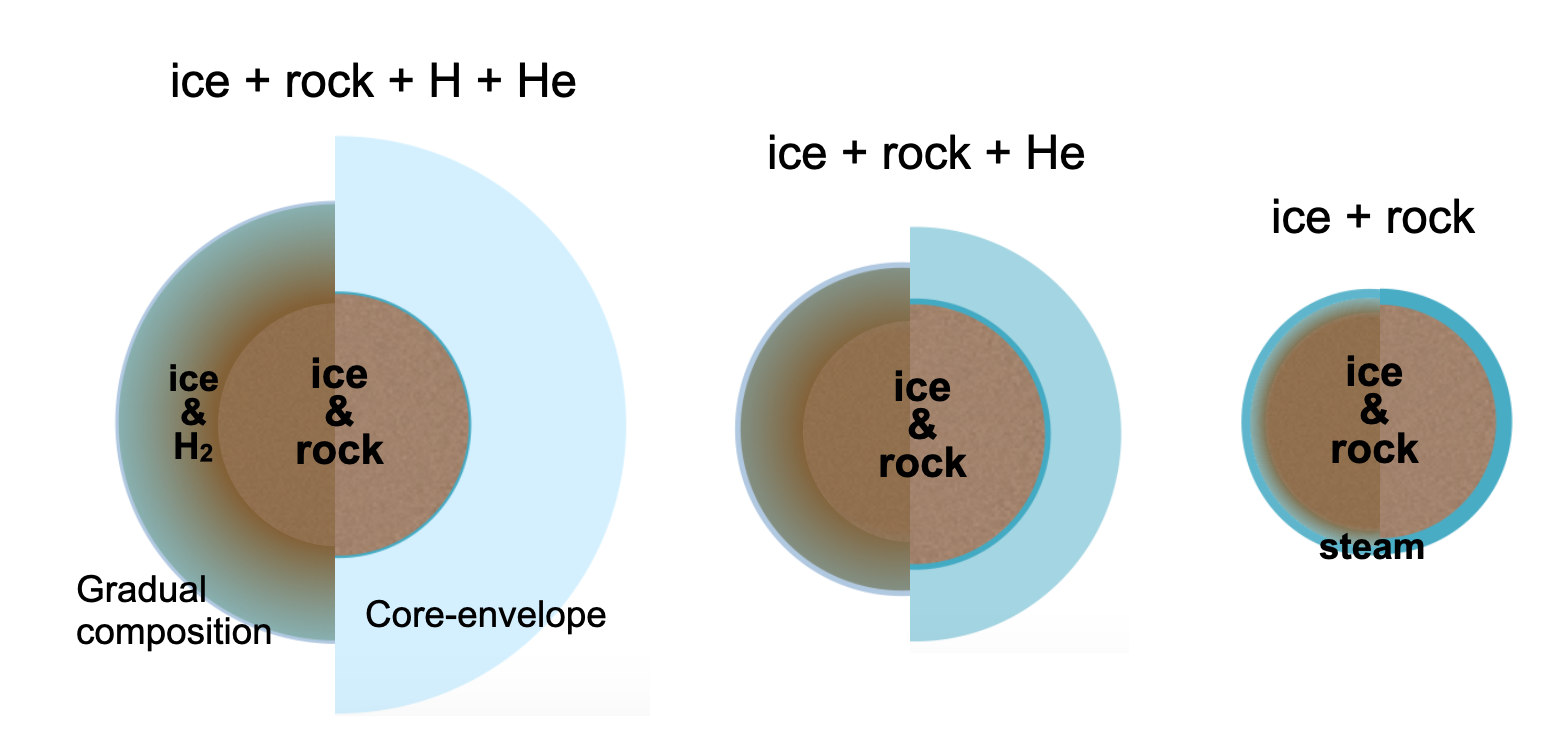}}
\caption{Schematic of the interior structure of ice-rich planets for different mass loss scenarios: no mass loss (left), hydrogen mass loss (middle), and all gas mass loss (right). Each sketch is split to show the two interior structures we examined. The ice \& rock mixture region covers more than 99\% of the planet mass for planets in the mass range of 5\me to 15\me (and above).}\label{fig:cartoon}
\end{figure}

We emphasize the interior structure for the 3 cases of mass loss and for the two interior structures in Fig.~\ref{fig:cartoon}. 
As all the gas is lost (right sketch), the two interior structures may become similar. The lower pressure in the outermost layers leads to ice-rock demixing. Consequently, the denser rock sinks to the mixed layer below, while the water forms an atmosphere around the mixed interior\footnote{The ice-rock separation can also happen when a small fraction of gas remains.}. For high stellar flux the volatiles are in the form of vapor, building a steam atmosphere\footnote{In highly irradiated envelopes with high water content the water molecules may dissociate and form an oxygen envelope, while the hydrogen escapes \citep{tian18}. Although this effect can be of great importance for the final planetary structure, we ignore it here for simplicity.} \citep{chambers17,turbet20}. 

\begin{figure}
\centerline{\includegraphics[width=9.5cm]{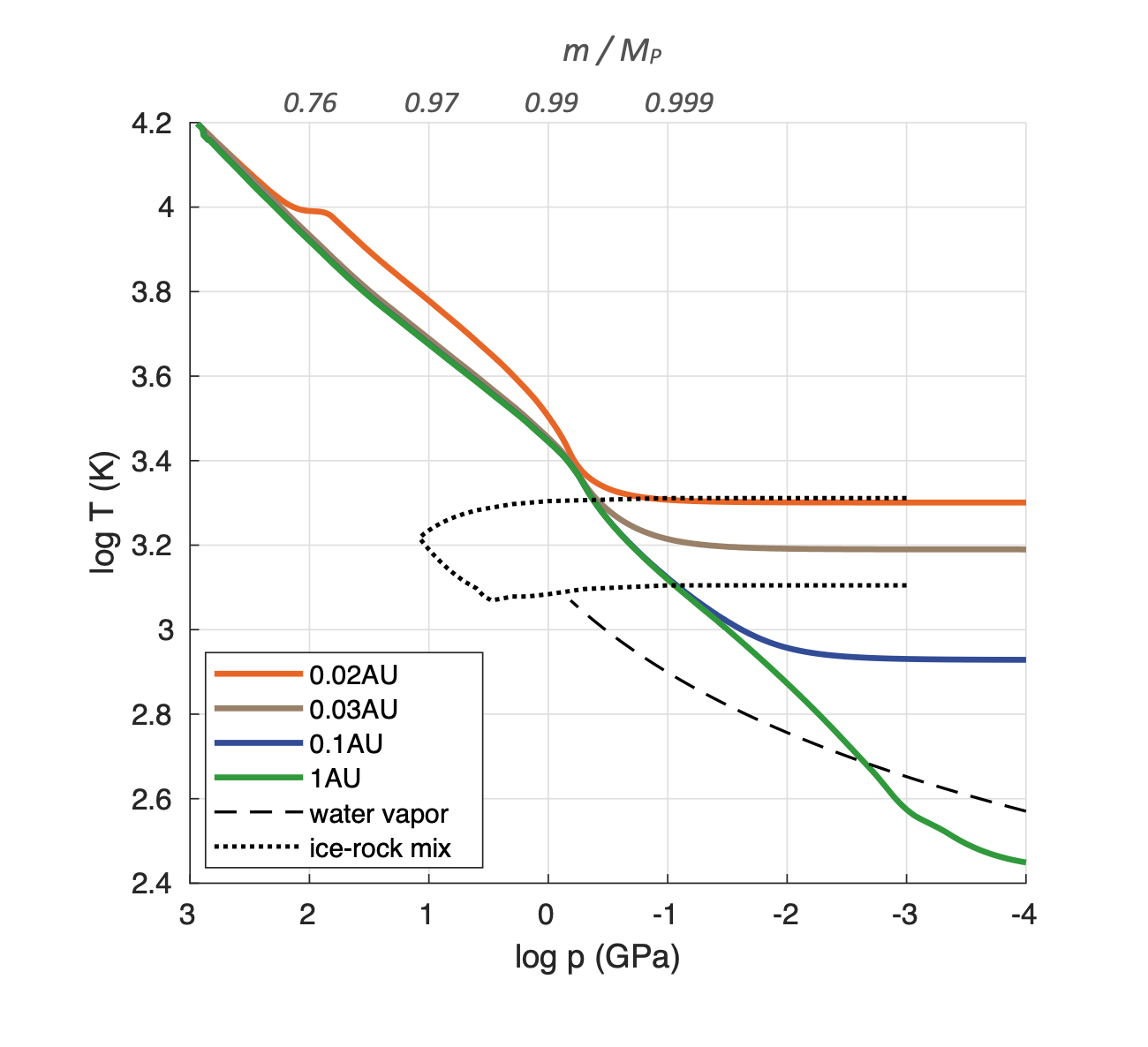}}
\caption{Pressure-temperature profiles for initially 15\me ice-rich planets that migrated to different distances from the star (solid colors), after mass loss. All planets have the same composition and structure as the brown curve of Fig.~\ref{fig:PT_005}. Curves of ice-rock interaction (dotted black) and water vapor (dashed black) are as in Fig.~\ref{fig:PT_space}. The upper axis signifies the mass fraction deeper than this pressure (similar for all cases). }\label{fig:PT_dist}
\end{figure}

Ice-rock tendency to demix on the surface (low pressure conditions) changes with temperature, and therefore is related to the distance from the parent star. 
In Fig.~\ref{fig:PT_dist} we show the temperature-pressure profiles of planets without gas envelope (similar to the brown curve in Fig.~\ref{fig:PT_005}) that have migrated to different distances from the star. 
The level of ice rock demixing on the planet's surface {facing the star} depends on the irradiation by the parent star: ice-rock are mixed up to the surface at 0.02AU (red), ice and rock are partially mixed on the surface at 0.03AU (gold), steam (water vapor) atmosphere is formed at 0.1AU (blue), and partially condensed water atmosphere at 1AU (green). 
The extremely high irradiation flux at 0.02AU causes an distinguishable increase in interior temperature by the slower interior cooling, in contrast to the further out interiors that we find to be similar. This increase {is partially} an overestimation by our 1D model that cannot properly simulate day-night effects{, which become important at very short  distances from the star}. Nonetheless, we limit our conclusion on similar interior properties for different distances from the star to be valid down to 0.03AU for our model parameters.

We suggest here, based on the nature of ice-rock interaction in planetary interiors, that {the mass fraction of the water surface shell is determined by the rock-ice tendency to demix}, as a function of the planet pressure (planet mass and gas mass) and temperature (thermal evolution and stellar flux), and {\it not} solely by the total ice fraction in the planet.

\subsection{Interiors of lower mass planets}\label{sec:lowm}

\begin{figure}
\centerline{\includegraphics[width=9.5cm]{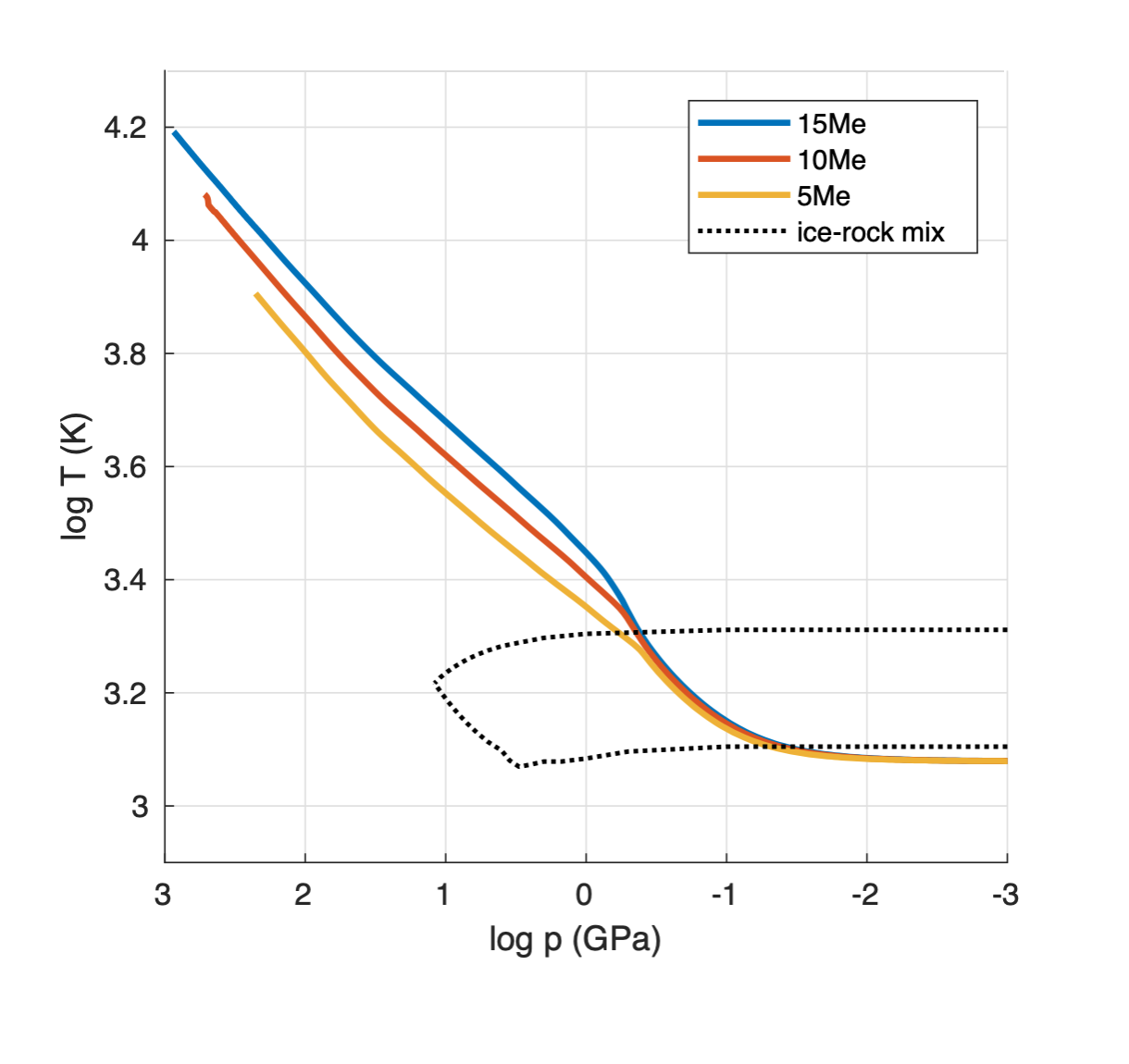}}
\caption{Pressure-temperature profiles for ice-rich planets at 0.05AU after gas loss for 3 initial masses of 15, 10, 5\me (blue, red, and yellow respectively). Ice-rock mix curve is as in Fig.~\ref{fig:PT_space}. For all cases, more than 99\% of the planet's mass is above the upper dotted curve (mixed), and less than 0.01\% of the mass is below the lower dotted curve (completely separated). }\label{fig:PTM_loss}
\end{figure}

We repeated the calculations shown in Fig.~\ref{fig:PT_005} for planets with lower initial masses and the same composition as the 15\me model. In Fig.~\ref{fig:PTM_loss} we show the pressure-temperature profiles for initially 5, 10, and 15\me planets after complete gas loss, at 0.05 AU. As expected, the mass fraction of the demixed ice and rock on the surface slightly increases with the decrease of planet mass. Lower mass planets have lower gravity and lower temperature and therefore have a larger fraction (mass fraction and absolute fraction) of separated ice {on the surface}. Nevertheless, for all cases ice and rock are miscible {in each other below the upper 1\% of the planetary mass (below 0.7\% for a 5\me planet, 0.2\% for a 10\me, and 0.1\% for a 15\me planet).}
Planets with no mass loss (not shown here) have larger fraction of mixed (miscible) ice and rock, as the fraction increases with the envelope mass, due to the increase in pressure in the metal-rich region. 
As ice-rich planets are predicted in the mass range of a few Earth masses and above \citep[e.g.,][]{venturini20b}, we suggest that ice and rock are miscible in each other in a significant fraction of the mass in ice-rich planets. 

\section{Discussion}\label{sec:dsc}

We find that interiors of close-in planets are similar in their properties to interiors of further out planets, down to 0.03AU for our model parameters. 
As a result, information from observations of close-in planets may tell us about their further-out twins. For example, measurements of obliquity and Love numbers of close-in planets \citep[e.g.,][]{kellermann18} might be useful to represent similar further-out planets. 
In the other direction, studying our solar system ice giants can help our understanding of close-in ice-rich exoplanets\footnote{In this context, in \cite{vazanhel20} we showed that Uranus models with gradual composition distribution and mixed ice and rock (1:2 and 2:1 mass ratios) can fit the measurements of Uranus.}. 

The radius valley in the Kepler's data \citep{fulton17} may be explained by two populations of dry and ice-rich planets \citep{zeng19,venturini20b}.
Observation of close-in planetary atmospheres in the coming years, by the ARIEL \citep{tinetti18} and JWST \citep{greene16} missions, will be able to shed some light on this topic by detecting the volatile abundance of planetary envelopes, and contribute the link of interior models to observations \citep{helled21}. 
Here, we suggest that the mass of the water/steam envelope of ice-rich planets is limited by the ice-rock interaction. 
Hence, 
the estimated mass of the envelope and its water content may not be representative of the bulk volatile content, as some of the ice may be locked in the rock in the deep interior.

We find in this study that ice and rock tend to mix in the interiors of 5-15\me planets, both for interiors with gradual composition distribution and for core-envelope structure.
This finding emphasizes the importance of material properties in determining the interior structure of planets. In order to properly determine the planetary structure the interactions between different species should be considered.
Material distribution in the interior affects trends in observation data, such as the planetary radius-mass relation {\citep{shah21}}, and correlation between planetary mass and envelope metallicity \citep{thorngren16,welbanks19,chachan19}.
Therefore, we suggest {that the mixed ice-rock interiors that} we present here {should be considered} when interpreting exoplanet observation data.

\subsection{Model caveats}\label{sec:dsc_cvt}

There are several caveats to our model. The first is the nature of the ice-rock interaction at pressure much higher than the SCP. 
We base our conclusions on the knowledge that in mixtures above the SCP ice and rock are in a supercritical phase and therefore completely miscible.
This behavior is well examined in lab experiments for pressure of a few GPa, and recently also for pressure of tens of GPa \citep{nisr20,kim21}. 
However, this behavior becomes unknown as we get deeper into the interior, toward pressures that are much higher, where experimental data isn't available.
Although we are not aware of any physical or chemical mechanism that is in favor of demixing at such high pressure, we cannot exclude molecular changes that may affect the miscibility.
{Rock} molecule dissociation \citep{melosh07}, and physical properties related to solid state phases at high pressure \citep[e.g.,][for ice and rock respectively]{redmer11,musella19} may affect our results and require further study.

The second caveat to our model is the existence of other materials in the mixture, rather than ice and rock.
Planetary materials usually do not occur as pure substances but as mixtures, where rocks belong to the MgO-\sio2-FeO complex and ices to the H-C-N-O mixtures, which have both a rich phase diagram. This may lead to new phenomena at high pressures such as demixing, known as helium rain in gas giants \citep[e.g.,][]{morales09} or carbon (diamond) rain in ice giants \citep[e.g.,][]{kraus17}.
Additional thermo-chemical processes, such as hydrogen mixing with molten rock \citep{hirschmann12b,chachan18,kite19}, iron \citep{stevenson77}, and volatiles  \citep{horiikoma11,kite20,bergermann21}, and interaction within the volatile family \citep{bethkenhagen17,nadenrobin18,guarguaglini19,nadenrobin20} can influence the pressure-temperature scheme at high pressure, toward more complex interior structure.

Another important caveat is the initial distribution of ice and rock in the interior.
Our results indicate that wherever rock and ice are mixed after planet formation, the ice-rock mixture is expected to remain mixed for the long term. Therefore, important factor for the ice-rock distribution in the interior is their initial distribution. Here, we assume accretion of mixed ice and rock, but if, for example, ice is being accreted after the rock then no significant mixing is expected. 
In the solid accretion process the size and accretion rate of solids and their volatility determine their deposition location and therefore the initial metal distribution. For example, icy pebbles evaporate exterior to rocky pebbles, while planetesimals of mixed ice and rock can get deeper as a mixture. 
In practice, the outer layers may be more ice-rich while the inner layers may be more rock-rich, with a gradual change in ice-rock mass ratio between them.
Our simplified model doesn't intend to model this, but to emphasize the importance and relevance of ice-rock interaction in the planet formation phase for the long term interior structure of the planet.

It should also be noted that our interior evolution model is 1D model. Therefore, 2D-3D effects such as orbital eccentricity of close-in Neptune planets \citep{correia20} and dayside to nightside differences are ignored. These simplifications can be justified, as the irradiation effect on the deep interior is found to be small for planets at a distance beyond 0.03 AU. 
We also ignore for simplicity effects of tidal heating \citep{henning09}, which may heat the interior further and increase the ice-rock miscibility, and of self-rotation, which under certain conditions limits the mixing of elements \citep{lock17}.

\subsection{Parameters' uncertainty}\label{sec:dsc_unc}

(1) The location of the ice-rock SCP can vary significantly between different laboratories and their experiments.
For example, the location of the wet solidus of the water - peridotitic rock system is highly debated 
\citep[e.g.,][]{kushiro68,millhoren74,mysen75,grove06,mibe07,green10,till12,kessel15,wang20}.
A summary of the experimental data of these works appears in Appendix~\ref{apx:curve} (Fig.~\ref{fig:solidi} and Table~\ref{tab:solidi}).
As can be seen from the data, 
the suggested temperatures for the wet solidus can differ by 200-600 degrees over the pressure range up to 6\,GPa. Only a few have attempted to determine the location of the second critical end point of the this mixture, where \cite{mibe07} and \cite{wang20} have suggested that the SCP is located between 3 and 4\,GPa, while \cite{melekhova07} have identified the SCP at 11\,GPa. 
Here, we adopt the higher value given in \cite{melekhova07}, i.e., the most conservative approximation for where ice and rock become completely miscible. Using the other values, or alternatively other rock types (see appendix~\ref{apx:rock}) will result in a larger fraction of the planet being in a mixed ice-rock phase.

(2) We assume an ice to rock {mass} ratio of 1:1 for formation exterior to the water iceline. However, water ice content exterior to the iceline can vary with stellar metallicity between 50\% and 6\% \citep{bitsch20}, and with short-lived radiogenic heating \citep{lichtenberg19}.
Atmosphere-free objects in the outer solar system have an ice to rock ratio of 1:2 \citep{nimmo19}. The lower (than 1:1) ice fraction of atmosphere free bodies may be the outcome of ice evaporation during violent solid accretion. 
A gaseous envelope may prevent the ices from escaping to space, as water-rich envelopes are less probable to escape \citep{biersteker21,aguichine21}.
The ice to rock mass ratio affects the mixing properties below the ice-rock SCP, as described in appendix~\ref{apx:curve}. Fortunately, the effect of the ice to rock ratio on the results we show here is limited; varying the water fraction between 5\% and 95\% shifts the miscibility threshold by about 200\,K \citep{kessel15}. On the wide temperature range in the interior of super-Earth to Neptune mass planets such a difference is insignificant for the point we make here. 

(3) The EoS of ice, rock, hydrogen, and helium, as well as the radiative opacity we use affect the results. New EoSs of water \citep{mazevet19} and of H, He \citep{chabrier19} exhibit higher densities than previous calculations, including this study. 
An increase in density is not expected to qualitatively change our findings.
{Moreover,} opacity of metal-rich envelopes, although crucial for the thermal evolution, can be strongly affected by the metals micro-physics in time \citep{movsh08} which is somewhat uncertain. Therefore, we tested several other values of higher and lower opacity. As expected, higher opacity slows the cooling and keeps the interior temperatures higher than is shown here, and lower opacity lead to slightly lower temperatures. In all the cases that we explored, the interior structure properties that we find here and the trends for migrated planets are similar.

(4) The initial gas fraction is another uncertain parameter that varies with formation model, mass loss at disk dissipation, giant impacts, and mass loss by stellar flux.
Nevertheless, a higher or lower gas fraction doesn't change our conclusions for the ice-rock interaction and the derived interior structure. 
Increasing the gas fraction of the envelope would increase pressure and temperature in the deep interior, and thus enlarge the {region} in which ice and rock are {chemically} mixed (miscible). Decreasing the gas mass fraction is modeled within the different mass loss cases shown in Fig.~\ref{fig:PT_005} and Fig.~\ref{fig:PTM_loss}.\\

Our study extends the variety of possible (standard) interior and evolution models for planets in the mass range of 5-15\me. 
The pressure range in the planets we study is up to about 10$^3$\,GPa, and the temperature varies up to several 10$^4$\,K. The above-mentioned points in this section emphasize the uncertainties we have today in our modeling, due to lack of knowledge of materials properties in this pressure-temperature regime. 
Future studies, experimental and computational, are encouraged in order to improve our understanding of the EoS and the phase diagram of planetary materials. Of particular interest for our models are: (1) the solubility and miscibility of elements in various multi-material systems. The miscibility, or immiscibility (demixing) is an important condition for possible differentiation processes in interiors; (2) the conductivity and viscosity of mixtures under high pressure. As these parameters have a significant effect on conduction and convection in the interior, they play a crucial role in the cooling of the planet.

\section{Conclusions}\label{sec:cncl}
\begin{enumerate}
\item Ice and rock tend to stay well mixed (miscible) in  interiors of ice-rich planet in the mass range of 5-15\me even after gigayears of thermal evolution, for the composition (EoS) we use in this work. 
\item Effect of migration on the thermal evolution of the deep interior is small. Deep interior structure and temperature of twin planets that migrated to different locations (beyond 0.03AU) are similar, if mass loss is insignificant. 
\item Mixing of elements by convection (convective-mixing) is limited in ice-rich planets. The gradual structure is mostly stable along the thermal evolution. Planets with gradual metal distribution have hotter deep interiors, in comparison to core-envelope structure. 
\item Ice-rich planets with substantial gas envelopes develop  ice-rock demixing only in the outer gas envelope, where water is miscible in the hydrogen.
\item Mass loss increases the metal enrichment of the envelope. For planets with gradual composition distribution the mass loss flattens the composition gradient, resulting in large scale convection and fast cooling.
\item In the absence of a hydrogen-helium envelope the ice and rock demix at the surface (at distances larger than 0.02AU). The rock is then dissolved in the deeper layers where ice and rock remain mixed, and the outer envelope is composed of {water} in steam / condensed form.
\item The total ice / water content of a planet cannot be inferred from the atmospheric mass and abundance, as a large fraction of the ice may be stored in the interior, mixed with the rock. 
\end{enumerate}

\section*{acknowledgements}
We thank the referee for useful comments. We thank Dave Stevenson and Morris Podolak for helpful discussions. We also thank Edwin Kite, Leslie Rogers, and Tim Lichtenberg for discussion during the Exoplanet-3 virtual meeting. AV acknowledges support by ISF grants 770/21 and 773/21. RS acknowledges the  supported of an ISF grant.

\bibliography{allona.bib} 
\bibliographystyle{aasjournal}

\appendix

\section{Critical points of ice and ice-rock systems}\label{apx:scp}

The \h2o fluid properties depend on its density, ordering, and hydrogen bonding and dissociation of the molecule. With increasing pressure and temperature the short-range ordered tetrahedral packing of \h2o molecules begins to break down, leading to largely disordered supercritical fluid, and the hydrogen-bond network is disrupted. This is the first critical end-point (also called the lower critical point), which refers to the one-component \h2o-only system. In \h2o-only system, the first critical end-point refers to the pressure-temperature conditions above which liquid water and vapor water are miscible, that is the boundary curve distinguishing between the stability field of water and the stability field of vapor terminates. Above these conditions, there is only one phase, a supercritical fluid. The water critical end-point is found at a temperature of 647\,K and a pressure of 22\,MPa.

Such behavior is also known in any rock-\h2o system, where \h2o-rich melts and rock-rich fluids become miscible in each other at high pressure-temperature. The physical and chemical properties of these two phases approach each other with increasing pressure and temperature. The second critical end-point (SCP, also referred to as the upper critical end-point) is a general term for all rock-\h2o systems.
Above this point the ice-rock system is in one supercritical fluid phase, while below it the cooling induces phase separation.

\section{Ice-rock interaction curves}\label{apx:curve}

\begin{figure*}
\subfigure{\includegraphics[width=8.8cm]{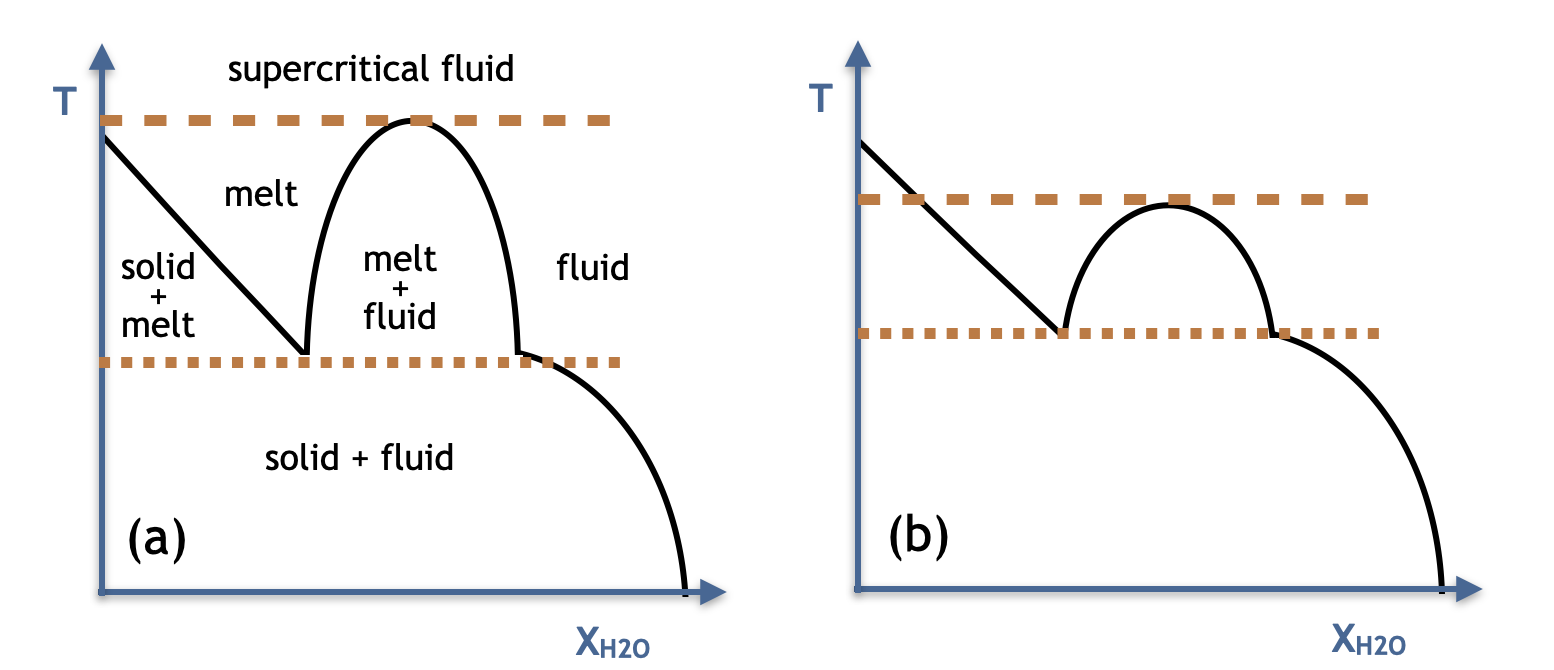}}
\subfigure{\includegraphics[width=8.8cm]{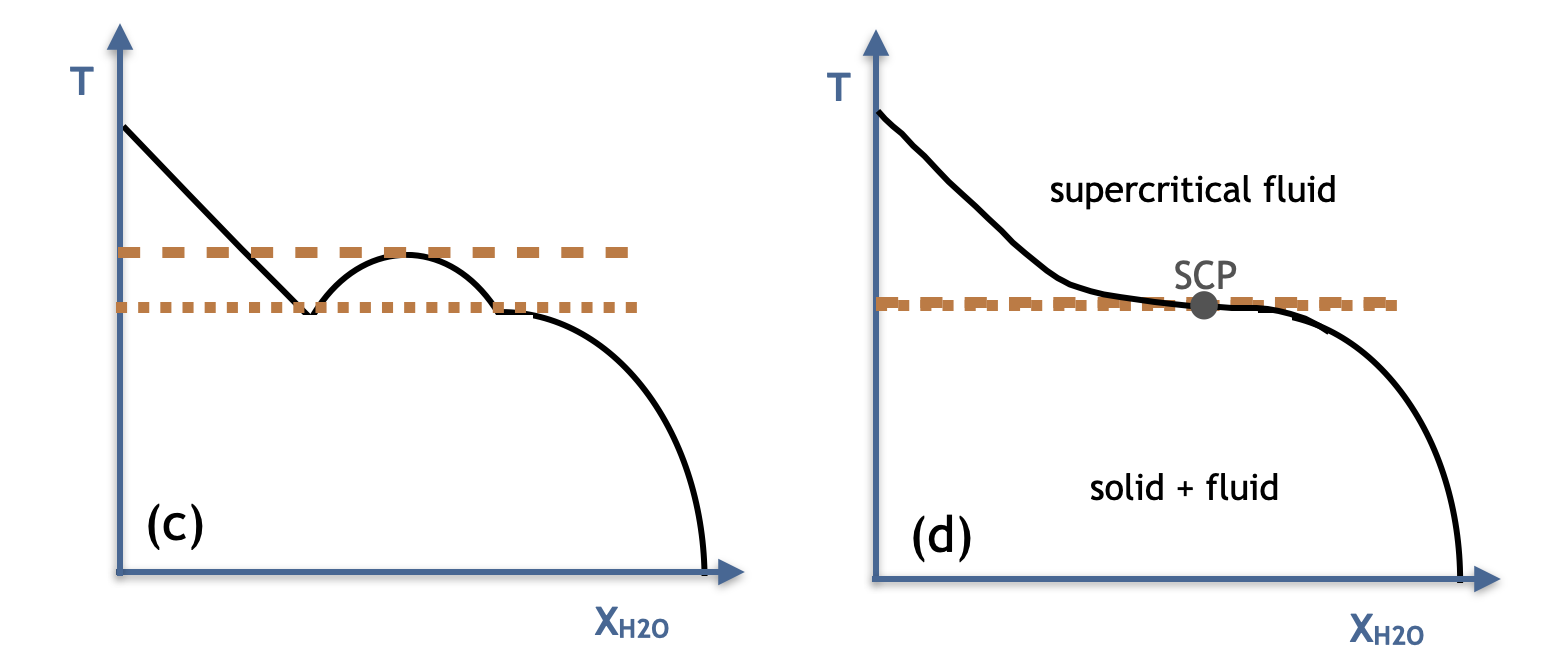}}
\caption{Schematic picture of ice-rock interaction as a function of the ice mass fraction (X$_{H_2O}$) and temperature (T). Each panel is a scheme in a constant pressure, where pressure increases from panels (a)-(d). The labels of the phases in panels (b) and (c) are similar to panel (a) and are omitted for clarity of the scheme. 
The characteristic temperatures of wet solidus (dotted curve) and miscibility (dashed curve) in each pressure are used to build the pressure temperature wet solidus and miscibility curves in Fig.~\ref{fig:PT_space}. As the miscibility temperature decreases with pressure the melt+fluid space (the parabola area) shrinks.  In panel (d) the mixture reaches the SCP pressure, where the wet solidus temperature (dotted) is equal the miscibility temperature (dashed).}\label{fig:Xice}
\end{figure*}

The ice-rock interaction is usually described in the temperature (T) - ice mass fraction (X$_{H_2O}$) space for a given pressure \citep[e.g.,][]{manning04}. In this space the different phases of the ice-rock system are shown. In Fig.~\ref{fig:Xice} we show schematic pictures of the equilibrium states for water-rock systems, with each panel representing a different pressure. At low pressure (left panel) and low temperature the rock is in solid form and in equilibrium with a fluid phase rich in water. As the temperature increases the ice-rock miscibility increases. At high enough temperature the ice-rock system becomes mixed into one supercritical fluid. 

The different panels show the change in the ice-rock system behavior with increasing pressure from left to right, starting from low pressure (a) up to the SCP point (d). As the pressure increases, the two-fluid (melt+fluid) field shrinks, up to the point where it completely disappears, which defines the SCP point \citep[see also Fig.~1 in][for 3D scheme]{stalder00}. The dotted orange curve signifies the wet solidus temperature for the given pressure, and the dashed orange curve is our estimate for a miscibility temperature. Below the wet solidus ice and rock are separated. Above the miscibility curve, ice and molten rock are completely miscible (mixed). In between these two temperatures the ice and rock are partially mixed in two phases: molten rock enriched with water, and water-rich fluid enriched with rock.
The difference between the wet solidus and the miscibility temperatures decreases with increasing pressure, up to the SCP where they coincide.

The wet solidus and miscibility temperatures for a given pressure, and their change with pressure, allow us to draw the wet solidus and miscibility curves in Fig.~\ref{fig:PT_space}. The wet solidus curve in Fig.~\ref{fig:PT_space} (brown) represents the collection of the dotted line values, and the miscibility curve (blue) represents the collection of the dashed line values. The values that we use in Fig.~\ref{fig:PT_space} are taken from \cite{kessel15} between 4-6\,GPa, \cite{grove06} between 1-3\,GPa, and \cite{melekhova07} between 11-13.5\,GPa.

It should be noted that there is no complete non schematic diagram for water-rock system. The shape of the melt + fluid region (the half ellipse in Fig.~\ref{fig:Xice}, the solvus) is not well constrained by the current experimental data. 
The wet solidus is directly measured in the experiment, but the miscibility curve is determined by theoretical knowledge of the diagram. One reason for that is that most of the laboratory experiments are motivated by the Earth studies, and therefore focus on a certain parameter range. The results for 50\% \h2o that we use in this work are therefore uncertain, where temperature uncertainties are usually on the order of tens of Kelvins \citep[see][]{kessel15}. More laboratory experiments are required in the high ice content regime, to study ice-rich interiors in more detail. 

{As is shown in Fig.~\ref{fig:Xice}, changing the \h2o mass fraction mainly affects the ice-rock interaction in the intermediate partially mixed region, between the solidus temperature (dotted) and the miscibility temperature (dashed). Above the miscibility temperature and below the solidus temperature the ice to rock mass ratio has no effect on the ice-rock interaction.}

\subsection{Experimental data for peridotite-\h2o systems}

\begin{figure}
\centerline{\includegraphics[width=9.5cm]{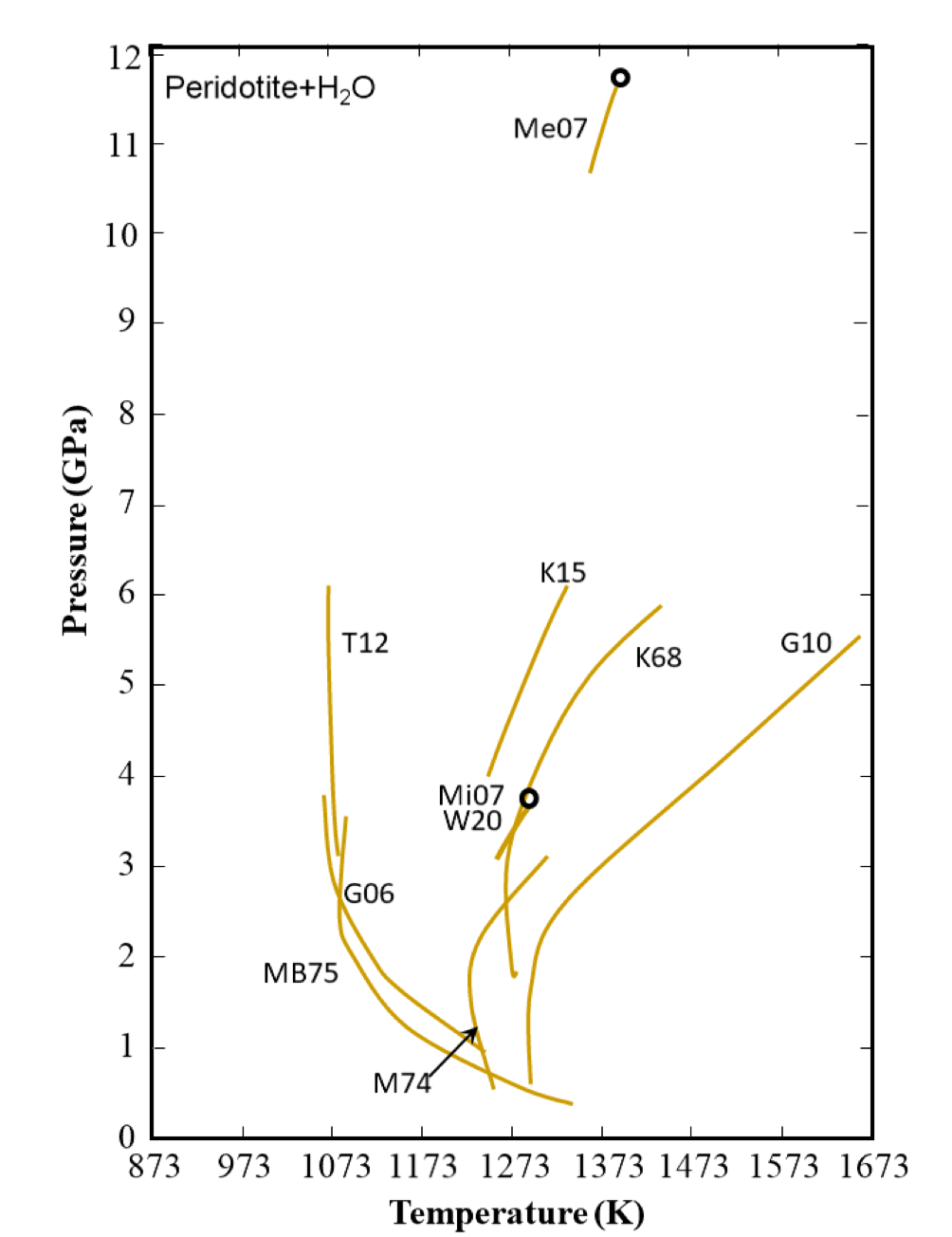}}
\caption{Literature compilation of solidi in the peridotite-\h2o systems: K68 = Kushiro et al. (1968); G10 = Green et al. (2010); M74 = Millhollen et al. (1974); MB75 = Mysen and Boettcher (1975); G06 = Grove et al. (2006); Me07 = Melekhova et al. (2007); Mi07 = Mibe et al. (2007); G10 = Green et al. (2010); T12 = Till et al. (2012); K15 = Kessel et al. (2015); W20 = Wang et al. (2020). Second critical endpoints (SCP) are marked with circle.}\label{fig:solidi}
\end{figure}

In Fig.~\ref{fig:solidi} and Table~\ref{tab:solidi} we summarize results from geophysical laboratory experiments at high pressure of peridotite rock with water.

\begin{table}
    \centering
    \begin{tabular}{|c|c|c|c|c|c|}
       \hline 
       Reference & p [GPa] &  T [K] & composition & wt\% \h2o & comments\\
       \hline \hline
       Kessel et al. (2015) & 4-6 & 1073-1473 & \shortstack{\sio2 Al$_2O_3$ Cr$_2O_3$ TiO$_2$ FeO MgO \\ CaO Na$_2$O K$_2$O \h2o} & 15 & \\
       \hline
       Melekhova et al. (2007) & 11-13.5 & 1273-2623 & \sio2 MgO \h2o  & 23 & SCP between 11-13.5 GPa \\
       \hline
       Grove et al. (2006) & 1-3 & 1053-1873 & \shortstack{\sio2 Al$_2O_3$ Cr$_2O_3$ TiO$_2$ FeO MgO \\ MnO NiO CaO Na$_2$O K$_2$O \h2o} & 14.5 & \\
       \hline
       Kushiro et al. (1968) & 0-5 & 1223-1873 & \shortstack{\sio2 Al$_2O_3$ Cr$_2O_3$ TiO$_2$ FeO MgO \\ MnO CaO Na$_2$O K$_2$O P$_2O_5$ \h2o} & 1 & \\
       \hline
       Millhollen et al. (1974) & 1-4 & 1223-1823 & \shortstack{\sio2 Al$_2O_3$ Cr$_2O_3$ TiO$_2$ FeO MgO \\ MnO NiO CaO Na$_2$O K$_2$O P$_2O_5$ \h2o} & 6 & \\
       \hline
       Mysen \& Boettcher (1975) & 0.5-3.5 & 1123-1523 & \shortstack{\sio2 Al$_2O_3$ Cr$_2O_3$ TiO$_2$ FeO MgO \\ MnO CaO Na$_2$O K$_2$O P$_2O_5$ \h2o} & 2 & \\
       \hline
       Mibe et al. (2007) & 1.7-4 & 1463-1623 & \shortstack{\sio2 Al$_2O_3$ FeO MgO \\ MnO Na$_2$O K$_2$O \h2o} & 60 & \shortstack{SCP between 3-4 GPa \\ Solidus was not determined} \\
       \hline
       Green et al. (2010) & 1.5-6 & 1223-1723 & \shortstack{\sio2 Al$_2O_3$ Cr$_2O_3$ TiO$_2$ FeO MgO \\ MnO NiO CaO Na$_2$O K$_2$O \h2o} & 1-2 & \\
       \hline
       Till et al. (2012) & 3.2-6 & 1013-1473 & \shortstack{\sio2 Al$_2O_3$ Cr$_2O_3$ TiO$_2$ FeO MgO \\ MnO NiO CaO Na$_2$O K$_2$O \h2o} & 14.5 & \\
       \hline
       Wang et al. (2020) & 3-6 & 1223--1473 & \shortstack{\sio2 Al$_2O_3$ TiO$_2$ FeO MgO \\ CaO Na$_2$O K$_2$O \h2o} & 10 & SCP between 3-4 GPa \\
       \hline
    \end{tabular}
    \caption{Composition, bulk water content, and second critical end-point (SCP), if exists, for the experimental peridotite-water systems presented in Fig.~\ref{fig:solidi}.}
    \label{tab:solidi}
\end{table}

\section{Rock type}\label{apx:rock}

In this study we take the peridotite as the best representative rock of the bulk planet composition in the solar system. 
The similarity in the abundance of non-volatile elements in the Sun and in some primitive meteorites (carbonaceous chondrite in composition) suggests that the compositions of all members of the solar system are related. The bulk composition of the Earth, as well as all other planets in the solar system, is considered chondritic. However, there are no available studies on the behavior of chondritic rock and water at high pressures and temperatures.
The Earth's mantle is the most voluminous layer of the Earth, divided into an upper and lower mantle. Upper mantle rocks exposed on the surface of the Earth have a peridotitic composition, which is only slightly depleted in Mg and Si compared to chondrites.

In other solar systems, and in the deep interior of the Earth, the type of rock may differ from the peridotite that we use as representative rock in this work.
The SCP varies with the type of the rock in the system.
The SCP of ice-rock system with pure silica ($SiO_2$) is around 1\,GPa \citep{kennedy62}, with granite rock it is at about 2\,GPa \citep{klimm08}, with basalt rock at 5.5\,GPa \citep{kessel05}, and with peridotite rock at 11\,GPa \citep{melekhova07}. 
These results suggest that the more refractory the composition is (i.e., lower Si and higher Mg content), the higher is the pressure of the SCP. 
Therefore, we believe that the SCP of a water-peridotite system lies at a pressure higher than 6\,GPa, and refers to the estimates appearing in this work as best resembling the interaction of water with the bulk of the Earth.

\end{document}